\documentstyle[12pt]{article}

\def\mat#1#2#3#4{{\left(\matrix{#1&#2\cr#3&#4\cr}\right)}}
\def\bM{{\bar M}}
\def\bA{{\bar A}}
\def\i{{\frac1{2\pi}\int d^2z\,}} 
\def\d{\partial}
\def\bd{\bar\partial}
\def\a{\alpha}
\def\b{\beta}
\def\n{\nabla}

\def\g{\gamma}
\def\bb{{\bar{b}}}
\def\by{{\bar{y}}}

\def\l{\lambda}
\def\s{\sigma}
\def\p{\phi}
\def\x{\chi}
\def\e{\epsilon}
\def\z{\zeta}
\def\za{\zeta_1}
\def\zb{\zeta_2}

\def\barg{{\bar g}}

\def\vb{\vec{\beta}}
\def\vx{\vec{\chi}}
\def\t{\tilde}
\def\implies{\Rightarrow}

\begin{document}

\newcommand{\inv}[1]{{#1}^{-1}} 

\renewcommand{\theequation}{\thesection.\arabic{equation}}
\newcommand{\beq}{\begin{equation}}
\newcommand{\eeq}[1]{\label{#1}\end{equation}}
\newcommand{\ber}{\begin{eqnarray}}
\newcommand{\eer}[1]{\label{#1}\end{eqnarray}}
\begin{center}
       August, 1994
                                \hfill    CERN-TH.7414/94\\
                                \hfill    RI-9-94\\
                                \hfill    WIS-7-94\\
                                \hfill    hep-th/9409011\\

\vskip .3in

{\large \bf Remarks on Non-Abelian Duality }
\vskip .4in

{\bf S. Elitzur}, \footnotemark\
\footnotetext{e-mail address: elitzur@vms.huji.ac.il}
{\bf A. Giveon}, \footnotemark\
\footnotetext{e-mail address: giveon@vms.huji.ac.il}
{\bf E. Rabinovici} \footnotemark\
\footnotetext{e-mail address: eliezer@vms.huji.ac.il}
\vskip .1in

{\em Racah Institute of Physics, The Hebrew University\\
  Jerusalem, 91904, Israel} \\

\vskip .15in

{\bf A. Schwimmer} \footnotemark \\

\footnotetext{e-mail address: ftschwim@weizmann.weizmann.ac.il}

\vskip .1in

{\em Weizmann Institute, Rehovot, Israel\\
and SISSA and INFN Trieste, Italy}

\vskip .15in

{\bf G. Veneziano} \footnotemark \\

\footnotetext{e-mail address: VENEZIA@NXTH04.CERN.CH}

\vskip .1in

{\em Theory Division CERN, 1211 Geneva 23, Switzerland\\
and Observatoire de Paris and ENS, Paris, France} \\

\vskip .1in
\end{center}
\vskip .2in
\begin{center} {\bf ABSTRACT } \end{center}
\begin{quotation}\noindent

A class of
two-dimensional globally scale-invariant, but not conformally
invariant, theories is obtained. These systems are identified in the
process of discussing global and local scaling properties of models related
by duality transformations, based on non-semisimple isometry groups.
The construction of the dual partner of a given model is followed through;
non-local as well as local versions of the former are discussed.

\end{quotation}
\vfill
\eject
\def\baselinestretch{1.2}
\baselineskip 16 pt
\noindent
\section{Introduction and Discussion}
\setcounter{equation}{0}

The large amount of constraints imposed by conformal invariance in
two dimensions has enabled the exact solution of many such systems.
The question has arisen time and again whether the same symmetry
and results could be obtained by requiring only the global scale
symmetry. In this paper we deal with various aspects of this issue
in two-dimensional field theories, aspects that have emerged rather
unexpectedly in the process of mapping one theory into another by
discrete transformations of target-space-duality type.

Duality relates two different geometries by establishing an isomorphism
between the sets of harmonic maps from $S^2$ into the two manifolds. A
standard procedure exists for discovering the dual partner
of a given manifold,
when the latter possesses a continuous group of isometries \cite{GPR}.

In ref. \cite{GRV}, a flat manifold was discovered for which the dual partner
resulting from this procedure, when applied
with respect to a certain group of
isometries, is a manifold whose corresponding two-dimensional sigma model
is not conformal. An explicit calculation of the $\b$-functions
corresponding to this model shows not only that they do not vanish,
but also that
they cannot be cancelled by an appropriate dilaton term. This
shows that duality, in the above sense of correspondence between classical
sigma model solutions, is not sufficient
for the equivalence of the two quantum
models on a curved worldsheet background.

It was pointed out in ref. \cite{GR} that the group of isometries used in
this example is non-semisimple, and contains  traceful structure constants.
It was also suggested that these features are related to an anomaly.
The anomaly was identified in ref. \cite{AAL}
as a mixed gravitational-gauge or conformal-gauge anomaly. Its origin is
in the possible dependence of the Jacobians -- related to the passage from
the original coordinates to the dual ones -- on the worldsheet metric
and the traceful isometry generators.

In this paper we show that the Jacobian appears in the form
$$\det (MN)/(\det M \det N),$$
and therefore, to have an overall anomaly it is crucial to have a
multiplicative anomaly \cite{K}, {\em i.e.},
$\det (MN)\neq \det M \det N$. It
turns out that, while  the dependence survives for the mixed
gauge-conformal anomaly, for the pure conformal anomaly it cancels.
As a result, in the dual sigma-model action the duality process generates
an additional non-local term proportional to the trace of the generator and
to ${1\over \Box}R^{(2)}$, where $R^{(2)}$ is the worldsheet curvature.
(This term is essentially the additional term which
duality induces from the anomaly term of \cite{AAL}.)

We use these results to study the worldsheet-metric dependence of models
emerging from such an anomalous duality process. Let A be the original
model with the invariance group of isometries, and B' the sigma model
resulting from the duality procedure (the ``prime" on B will be explained
later). We first stress that, since the
only difference between B' and A is the calculable and well-controlled
non-local anomalous term, the extra dependence of B' on the conformal
background should cancel exactly the background dependence of this term. In
particular, if A is conformal, the ordinary geometric calculation of the
$\b$-function of B' should be cancelled by the variation of the
anomaly term with respect to the background conformal factor. We check this
in detail for a few examples, including that studied in \cite{GRV}.

On a flat worldsheet background, the sigma models A and B' are already
equivalent. In particular, their spectra are identical. This is confirmed
by the result that the anomaly is proportional to $R^{(2)}$. In the case
in which A is conformal, the spectrum of B' must be massless as well.
Next we recall some general features of massless 2-$d$ sigma models.

The simplest situation is  when the model is truly conformal, {\em i.e.},
when there exists a local, traceless energy-momentum tensor $T_{\a\b}$.
The $\b$-function  equations  give a vanishing result. As a consequence,
in this case the model can be coupled to a two-dimensional background
metric in a Weyl-invariant manner at the quantum level.
It could happen that such a local traceless tensor  does not
exist. Since the spectrum is massless, there should exist a local, conserved
dilation current $D_{\a}$. From the relation
between  $T_{\a\b}$ and $D_{\a}$ it follows \cite{polch}
that in this case the trace of the energy-momentum tensor
must be a total derivative. The $\b$-function equations no longer
give  zero; however, the deviation from zero can be absorbed
in a ``wave function renormalization", {\em  i.e.}, a reparametrization
of the fields, and in a gauge transformation of the antisymmetric
tensor which enters the sigma model \cite{frompolch,polch}.
Explicitly, the $\b$-function
with respect to the target-space metric $G_{\mu\nu}$ has the form:
\beq
\b_{G_{\mu\nu}}=\n_{\mu}\xi_{\nu}+\n_{\nu}\xi_{\mu},
\eeq{bxx}
where $\xi_{\mu}$ defines the reparametrization. For a general
$\xi_{\mu}$ one cannot define a local Weyl-invariant coupling
to the background metric. If, however, $\xi_{\mu}$ has the special form:
\beq
\xi_{\mu}=\d_{\mu}\Phi,
\eeq{zeta}
where $\Phi$ is the ``dilaton" field, a local coupling can be defined
involving the curvature, and an ``improved" traceless energy-momentum
tensor exists.

Our case of a sigma model B', resulting by an anomalous duality process
from an originally conformal theory, is of the type of a massless
{\em nonconformal} theory. In fact,  the representation of its
$\b$-function as resulting from the variation of the anomalous term with
respect to the conformal worldsheet factor, makes it explicit that these
$\b$-functions are total derivatives  representable in the form of eq.
(\ref{bxx}).

We will study two such examples.
One will turn out to be improvable by a dilaton term. The other example,
that of \cite{GRV}, develops a $\b$-function which is a total derivative,
but not improvable by a local background-dependent term. We have thus found
a genuine example of a two-dimensional, scale-invariant, interacting
non-conformal model. In ref. \cite{polch} it was proved that for a unitary
theory in $d=2$ with a discrete operator spectrum, an improved
energy-momentum tensor always exists. Our example is, necessarily, a
non-compact sigma model. Similar situations arise in higher dimensions at
the tree level for gauge theories based on $n$-forms \cite{DS}.

Note, however, that this apparently non-locally
improvable B' model is actually improvable in a more general sense. One can
``improve" it by adding to its action the non-local background-dependent
anomaly term, and then re-express it in a local form
by passing to the language of the
A-model which, in the presence of the anomaly term, is completely
equivalent to it.

The paper is organized as follows: in section 2, we discuss in detail
the procedure of non-Abelian duality \cite{nonad}
(for an isometry group acting with no
fixed points \cite{GR}), in particular the appearance of the
anomaly. In section 3, we show how the original A-model can be recovered.
In section 4, we discuss the B-model related to it, {\em i.e}, the B'
sigma model together with the (non-local) improvement.
In Section 5, we discuss an example where the reparametrization
needed is of the form (\ref {zeta}) and, therefore, models A and B
are equivalent after a dilaton-type correction to B' is made.
In section 6, we discuss the model proposed in \cite{GRV}: model A is
conformal but model B' is just dilation invariant.
In section 7, we discuss how the anomaly can be understood in terms
of Ward identities in a flat worldsheet background.
In section 8, we discuss some alternative local representations of the mixed
anomaly term, in a higher dimensional sigma model.
A general form of the anomaly for a non-compact group and a general
proof of the dilaton corrections \cite{buscher} are discussed
in two appendices.

\section{The general case (without isotropy)}
\setcounter{equation}{0}

As in ref. \cite{GR}, we can consider a target space
with coordinates $g$ that transform as $g\to u g$ for $u$ in some
group $G$, and further coordinates $x^i$ that are inert.
A general action can be written in the form
\ber
S[g,x]=\i \Big( E_{ab}(x)(\inv g\d g)^a(\inv g\bd g)^b &+&
F^R_{aj}(x)(\inv g\d g)^a\bd x^j \nonumber\\
+\ \ F^L_{ib}(x)\d x^i(\inv g\bd g)^b &+&F_{ij}(x)\d x^i\bd
x^j \nonumber\\
&-&\Phi (x)\d\bd\s\Big)\ ,
\eer{noiso}
where
\beq
(\inv g\d g)^a \equiv tr (\t{T}^a \inv g\d g) \ \ \Leftrightarrow \ \ \inv
g\d g = (\inv g\d g)^a T_a\ ,\ \ etc.
\eeq{gdga}
The generators $T_a$, $a=1,...,dim(G)$, obey
\beq
[T_a,T_b]=f^c_{ab}T_c,
\eeq{TTfT}
and the ``dual generators'' $\t{T}^a$ are defined by the condition
\beq
tr(T_a \t{T}^b)= \delta_{a}^b.
\eeq{trTT}
In eq. (\ref{noiso}) the background matrices $E_{ab}, F^L_{ib}, F^R_{aj},
F_{ij}$ and the dilaton $\Phi$ depend only on the coordinates $x^i$.
Here $z=(\za+i\zb)/\sqrt{2}$, $\bar z = (\za-i\zb)/\sqrt{2}$
are complex worldsheet coordinates,
$\d\equiv \d/\d z=(\d_1-i\d_2)/\sqrt{2}$, $\bd\equiv \d/\d\bar
z=(\d_1+i\d_2)/\sqrt{2}$, and $\s(z,\bar z)$
is the worldsheet conformal factor, {\it i.e.}
\beq
\d\bd\s={1\over 4}\sqrt{h}R^{(2)},\qquad  h_{z\bar z}=e^{-2\s},
\eeq{Rs}
where $h$ is the worldsheet metric (in the conformal gauge) and $R^{(2)}$
is the worldsheet curvature.

The quantum field theory is defined by the functional integral
\beq
Z_O=\int D_Lg\ Dx\ e^{-S[g,x]}
\eeq {pint00}
(for future use, we label this partition function as $Z_O$).
Here $D_Lg$ is the left invariant measure (which is required for
consistency with the isometry that acts from the left, $g\to u
g$)~\footnote{
For semi-simple groups there is no difference between left and right
invariant measures; however, for general groups they may differ.
}
\beq
D_Lg\equiv \prod_{a,z,\bar z} (\inv g dg)^a,
\eeq{DLg}
and $Dx$ is the rest of the sigma-model measure ($Dx\equiv \prod_{z,\bar
z}\sqrt{\det(F_{ij})\det(E_{ab})}dx$, if $F^L=F^R=0$).

We now rewrite the theory (\ref{pint00}) by inserting the identity
$I=\int D_L\barg\ \delta_L(g,\barg)$. The (left-invariant) delta-function
sets $\bar{g}=g$ and allows, in particular, the replacement of
$\inv g \bd g$ by $\inv{\bar{g}} \bd \bar{g}$:
\beq
Z_O=\int D_Lg\ D_L\bar{g}\ \delta_L(g,\bar{g})\ Dx\ e^{-S[g,\bar{g},x]},
\eeq{pint000}
where
\ber
S[g,\bar{g},x] &=& \i \Big( E_{ab}(x)(\inv g\d g)^a(\inv \barg \bd \barg)^b\
+\ F^R_{aj}(x)(\inv g\d g)^a\bd x^j \nonumber\\
&+&\ F^L_{ib}(x)\d x^i(\inv \barg\bd \barg)^b\ +\ F_{ij}(x)\d x^i\bd
x^j\nonumber\\&-& \frac14\Phi(x)\sqrt{h}R^{(2)}\Big) \, .
\eer{Sxgg}

At this point we define a vector field  $A,\bA$ in terms of
the group variables $g,\bar{g}\in G$:
\beq
A=\inv g \d g, \qquad \bA=\inv{\barg}\bd \barg.
\eeq{AAgg}
Changing the integration variables from $g,\barg$ to $A,\bA$
one gets
\ber
Z_O \to Z &=& \int DA\ D\bA\
\det(DA/D_Lg)^{-1}\ \det(D\bA/D_L\barg)^{-1} \nonumber\\
&\times& \delta(F)\ \det(DF/D_Lg)|_{\barg=g}\
e^{-S[A,\bA,x]},
\eer{SxAA}
where $S[A,\bA,x]$ is given by inserting eq. (\ref{AAgg}) into
$S[g,\barg,x]$ (\ref{Sxgg}).
In eq. (\ref{SxAA}) $F$ is the field strength,
\ber
F(A,\bA)&=&\d\bA-\bd A+[A,\bA]\ =\ \inv g\d(\inv f\bd f)g\
=\ \inv{\barg}\bd(\d f\inv f)\barg, \nonumber\\
f&\equiv&\barg\inv g,
\eer{FAA}
and we have used the equality\footnote{
The equality (\ref{Fgg})
is only correct locally: $\delta(F)$ constrains the curvature
to vanish, but does not force the connection $A$ to be trivial on a
non-zero genus worldsheet. Therefore, to correct eq. (\ref{Fgg}), we
need to sum $\delta_L(g,\barg)$ over all flat connections. We will not
address such global issues \cite{RV,GR,AABL} here.
Moreover, as $F=0$, one could involve in the process of duality any
function of $F$ which would vanish for $F=0$. Such terms lead, in general,
to very complicated non-local theories, all equivalent. This structure
can, however, be removed by applying non-local field redefinition.}
\beq
\delta_L(g,\barg)=\delta(F)\det N, \qquad N=(DF/D_Lg) |_{\barg=g}.
\eeq{Fgg}
Calculating $N$ (by using  eqs. (\ref{DLg}) and (\ref{FAA}))
we obtain formally
\beq
N=M(A)\bM(\bA),
\eeq{detF}
where
\beq
M(A)=\d+[A,\cdot], \qquad \bM(\bA)=\bd+[\bA,\cdot].
\eeq{MbM}

Next we replace the $F=0$ delta-function with the constraint imposed
by a Lagrange multiplier term in the action, namely,
\beq
\delta(F)=\int D\l\ e^{-\i tr\l F} .
\eeq{dFDl}
Here
\beq
F\equiv F^aT_a,\quad \l\equiv \l_a\t{T}^a\ \implies \ tr\l F= \l_a F^a.
\eeq{FllF}
Finally, we should deal with the Jacobian of the change of variables from
$g,\barg$ to $A,\bA$:
\beq
\det(DA/D_Lg)\det(D\bA/D_L\barg)=\det M(A) \det \bM(\bA),
\eeq{detAh}
where $M(A),\bM(\bA)$ are given in (\ref{MbM}). The precise manner of
calculating these determinants is given in \cite{AAL}.

Altogether, the partition function takes the form:
\ber
Z&=&\int D\l\ Dx\ DA\ D\bA\ e^{J(A,\bA)}\ e^{-S[A,\bA,\l,x]},\nonumber\\
e^{J(A,\bA)}&=&{\det(M(A)\bM(\bA))\over \det M(A) \det \bM(\bA)},
\eer{ZlAA}
where
\ber
S[A,\bA,\l,x] &=& \i \Big( E_{ab}(x)A^a\bA^b\
+\ F^R_{aj}(x)A^a\bd x^j \nonumber\\
&+&\ F^L_{ib}(x)\d x^i\bA^b\ +\ F_{ij}(x)\d x^i\bd
x^j\ -\ \frac14\Phi(x)\sqrt{h}R^{(2)}\nonumber\\
&+&\l_a(\d\bA^a-\bd A^a +f^a_{bc} A^b\bA^c)
\Big) \ .
\eer{SxA}

In the following, we discuss the evaluation of the determinants term in
$Z$ (\ref{ZlAA}): $\exp J(A,\bA)$. Naively
the determinants would cancel; however, there is a possibility of a
multiplicative anomaly \cite{K}. It is convenient to use a covariant
notation and to replace the determinants by actions involving bosonic
ghosts $\b_{\a},\g,\g'$ (here $\a=1,2$ is a worldsheet index)
and fermionic ghosts $b,c$:
\ber
e^{J(A,\bA)}&=&\int D\b\ D\g\ D\g'\ Db\ Dc
\nonumber\\
&\times& \exp \int d^2\z\ tr[\sqrt{h}h^{\a\delta}
(\b_{\a}D_{\delta}\g+bD_{\a}D_{\delta}c)+\b_{\a}\t{D}^{\a}\g'],
\eer{bcbc}
where
\beq
\b_{\a}\equiv \b_{\a a}\t{T}^a,\qquad \g\equiv\g^aT_a,\qquad
\g'\equiv\g'^aT_a,\qquad b\equiv b_a\t{T}^a,\qquad c\equiv c^a T_a ,
\eeq{bggbc}
\beq
D_{\a}\equiv \d_{\a}+[\inv g \d_{\a} g, \cdot], \qquad \t{D}^{\a}\equiv
\e^{\a\b}D_{\b}.
\eeq{tD}
In deriving (\ref{bcbc}) we used the fact that
\beq
(D^{\a}+\t{D}^{\a})(D_{\a}-\t{D}_{\a})=2D^{\a}D_{\a}
\eeq{if}
if the field entering $D_{\a}$ is a pure gauge.

The ghost actions are formally conformal invariant as well as gauge
invariant under $g\to gu$, with all ghost fields
transforming in the adjoint representation:
$ c\to \inv u c u$, $b\to \inv u b u $. A new type of
anomaly may appear coupling the
background worldsheet metric to the gauge field  \cite{GR,AAL}. This
anomaly can be studied by considering for the free part of (\ref{bcbc})
the correlator between the energy-momentum tensor and the vector current
$V_a^{\a}$ coupled to the gauge field (note that by giving a negative
intrinsic parity to $\g'$, its gauge coupling can also be considered
vectorlike):
\beq
{\delta^2 J\over \sqrt{h}\delta h_{\a\b}(\z)
\sqrt{h}\delta A_{\g}^a(\xi)}=
\langle T^{\a\b}(\z) V_a^{\g}(\xi) \rangle \equiv S_a^{\a\b\g}(\z-\xi),
\eeq{TVS}
\ber
T^{\a\b}={1\over 2}\sqrt{h}\ tr\Big(\b^{\a}D^{\b}\g-{1\over
2}h^{\a\b}\b^{\delta}D_{\delta}\g\Big)+(\a\leftrightarrow\b),
\nonumber \\
V_a^{\a}\equiv {1\over\sqrt{h}}{\delta J\over \delta
A_{\a}^a}=f_{ad}^e(\b^{\a}_e\g^d+\e^{\delta\a}\b_{\delta e}\g'^d
+2b_e\n^{\a}c^d).
\eer{VdJ}
Note that $T^{\a\b}$ in (\ref{VdJ}) does not involve $\g'$.
This energy-momentum tensor is invariant under the group $G$. Taking
the expectation value of the commutator of $T^{\a\b}V^{\g}_a$ with an
isometry generator $Q_b$, and using $[Q_b,T^{\a\b}]=0$, and
$Q_b|0\rangle=0$ which implies
$\langle [Q_b,T^{\a\b}V^{\g}_a]\rangle=0$, one obtains
\beq
\langle T^{\a\b} [Q_b,V_a^{\g}] \rangle = 0.
\eeq{TQV}
Therefore, the correlator (\ref{TVS})
can be non-zero only for a group index which does
not appear on the right-hand side of a commutator $[Q_b,V_a]$. Note that
all the generators which do appear on this right-hand side necessarily have
zero trace (because the trace of a commutator is zero). The
necessary and sufficient condition to have such ``quasi-Abelian''
directions is for the group to be non-semisimple \cite{H}.

The kinematical constraints on the correlator (\ref{TVS}) on a flat
background are very simple to analyse:
the correlator depends on $\z-\xi$ and, therefore, its Fourier transform
$\t{S}$ depends on a single momentum $q^{\a}$, and its general
kinematical decomposition has the form
\beq
\t{S}^{\a\b\g}_a=A_a \eta^{\a\b}q^{\g} + B_a (\eta^{\a\g}q^{\b}+
\eta^{\b\g}q^{\a})+C_a q^{\a}q^{\b}q^{\g},
\eeq{KKK}
where $A_a,B_a,C_a$ are invariant amplitudes depending on $q^2$ only.
Using eq. (\ref{KKK}) it is easy to show that at least two out of the three
Ward identities, following from the conservation and tracelessness of
$T^{\a\b}$ and the conservation of $V_a^{\a}$, should be violated.
In particular,
if the conservation of $T^{\a\b}$ is imposed, $\t{S}^{\a\b\g}_a$ has a unique
form in any dimension $d$:
\beq
\t{S}_a^{\a\b\g}=K_a(\eta^{\a\b}q^2-q^{\a}q^{\b})(q^2)^{(d-4)/2}q^{\g},
\eeq{SKq}
where $K_a$ are dimensionless constants. The expression (\ref{SKq}) violates
the conservation of the vector current. It follows that, if dimensional
regularization can be used, the correlator will vanish identically.
Therefore, the only possibility to obtain an anomalous correlator is to
have in the action a $\g_5$ or $\e^{\a\b}$.
We conclude, therefore, that the $b,c$ action corresponding to
$\det(M(A)\bM(\bA))$ does not have an anomaly~\footnote{
If we studied $\det(M(A)\bM(\bA))$  in a general
$A$-background, the anomaly could appear only in the unnatural
parity part; for this type of anomaly the current conservation
can be imposed, and therefore the anomaly vanishes for a pure
gauge configuration, leading again to the same conclusion.}.
In order to calculate the coefficient $K$ we isolate the
convergent $C$ amplitude in the Feynman diagram, the other amplitudes
being determined by the choice of which Ward identity is preserved.
We obtain: $K_a=1/4\pi\ (tr\ T_a)$.

On a general background, covariantizing (\ref{SKq}) and using eq. (\ref{TVS}),
the simultaneous variation of $J$ with respect to the background metric
and gauge field is
\ber
\delta J&=& {1\over 4\pi} (tr\ T_a)\int d^2\z d^2\xi\ S_a^{\a\b\g}(\z-\xi)
\sqrt{h}\delta h_{\a\b}(\z)\sqrt{h}\delta A_{\g}^a(\xi)\nonumber\\
&=&{1\over 4\pi} (tr\ T_a) \int d^2\z d^2\xi
\sqrt{h(\z)}\sqrt{h(\xi)}\nonumber\\
&\times& \Big({h^{\a\b}\over\sqrt{h}}\delta^2(\z-\xi)-
{\n^{\a}\n^{\b}\over \Box}\Big)\delta \n^{\g} A_{\g}^a(\xi) \delta
h_{\a\b}(\z).
\eer{delJ}
In the conformal gauge, $h_{\a\b}=\exp(-2\s)\eta_{\a\b}$ (\ref{Rs}),
one finds
\beq
J=-{1\over 2\pi}(tr T_a) \int d^2\z \sqrt{h}h^{\a\b}\ \n_{\a}A_{\b}^a\ \s .
\eeq{JAs}

We remark that, potentially, there could be an anomalous contribution
involving just the metric (the standard Polyakov anomaly,
$\sqrt{h}R^{(2)}{1\over\Box}R^{(2)}$).
The explicit expression (\ref{bcbc}) shows that we have a bosonic $\b$-$\g$
system, with $c=2$, and an anticommuting complex scalar ($b$-$c$), with
$c=-2$, coupled to the background worldsheet metric. Therefore, the anomalies
cancel and there is no multiplicative anomaly depending just
on the metric.
The central charge, $c_t$, of the system described by eq. (\ref{ZlAA}) is
equal, by construction, to the system described by eq. (\ref{noiso}).
For semisimple groups, one can show that the system in eq. (\ref{SxA})
already has
a central charge $c_t$, leaving for the rest of the system,  $e^J$
in (\ref{ZlAA}), only a zero central charge. This result was obtained above
by assuming that $(b,c)$ was a scalar pair. With such a scalar pair, one
can show that the numerator in $e^J$ (\ref{ZlAA}) can be regularized in
such a manner that it leads to no mixed anomaly.

{}From eq. (\ref{JAs}), we see that
for a non-trivial worldsheet metric and gauge fields,
an anomalous (non-local) piece appears in the effective action:
\ber
S_{nonlocal}&=&{1\over 8\pi}(tr\ T_a)\int d^2z \Big({1\over \d}A^a+{1\over
\bd}\bA^a\Big)\sqrt{h}R^{(2)}\nonumber\\
&=&{1\over 2\pi}\int d^2z\ ln(\det M(g)\det M(\bar{g}))\d\bd\s,
\eer{As}
(here we are back to complex worldsheet coordinates)
where $M_a^{\ b}(g)$ is the adjoint representation matrix, {\it i.e.}
\beq
gT_a\inv g = M_a^{\ b}(g)T_b\ \implies \  M_a^{\ b}(g)=tr(gT_a\inv g
\t{T}^b).
\eeq{Madj}
The proof of the second equality in (\ref{As}) is given in Appendix A.
We should note here that $\det M$ is {\em not} a constant if $T_a$ is not
traceless for some $a$:
\beq
tr\ T_a\neq 0\ \Leftrightarrow \ \det\ M \neq 1 .
\eeq{detM1}
Explicitly, if we parametrize
\beq
g(\theta)=e^{\theta^a T_a},
\eeq{gthe}
then
\beq
ln\ \det\ g = tr\ ln\ g =  tr\ \theta^a T_a.
\eeq{lndetg}
(Obviously, the matrix $g$ is equivalent to $M(g)$ if the generators $T_a$
are in the adjoint representation.)

To summarize, the theory is now defined by the functional integral
\ber
Z\ =\ \int DA\ D\bA\ D\l\ Dx\ e^{-(S[A,\bA,\l, x]+\i
ln(\det M(g) \det M(\bar{g}))\d\bd\s)},\nonumber\\
\eer {pint}
where $S[A,\bA,\l,x]$ is given in (\ref{SxA}).

Starting with the theory (\ref{pint}),
we now consider two avenues which we call the ``A-model'' and the
``B-model''.
One gets the A-model by integrating out the Lagrange multiplier.
On the other hand, one gets the B-model by integrating out the gauge
field\footnote{
An alternative way \cite{RV} to get (\ref{pint}) from (\ref{pint00}) is
to gauge the $G$-symmetry of the action $S[g,x]$
with non-dynamical gauge fields ({\em i.e.}, without a $F^2$
term) by minimal coupling:
$$ \inv g\d g \to \inv g(\d +A)g\ ,\ \ \
\inv g\bd g \to \inv g(\bd +\bA)g,$$
(here $A\equiv A^aT_a$ transforms as $A\to u(\d +A)\inv u$, {\it etc.}),
and add the Lagrange multiplier term, $tr\l F$,
that constrains the gauge field to be (locally) pure gauge. Then, after
choosing a unitary gauge, $g=1$, and adding the non-local term to cancel the
trace anomaly \cite{GR,AAL}, one recovers (\ref{pint}).
We chose not to work with such a
gauge theory, in order not to deal with the problem of a gauge-invariant
measure for the gauge field in non-semisimple groups.}.

\section{The A-model}
\setcounter{equation}{0}

By construction, integrating out $\l$ in (\ref{pint}) constrains the gauge
field to be pure gauge ($g=\bar{g}$), and we get
\ber
Z_A&=&\int DA\ D\bA\ \delta(F)\ Dx\
e^{-(S[A,\bA,\l,x]+S_{nonlocal}[A,\bA,\s])}\nonumber\\
&=&\int D_Lg\ D_L\bar{g}\ \delta_L(g,\bar{g})\ Dx\ e^{-S[g,\bar{g}, x]}
\nonumber\\ &=&\int D_Lg\ Dx\ e^{-S[g,x]}.
\eer{modA}
Here $S[g,x]$ is given by (\ref{noiso}) and
$S_{nonlocal}[A,\bA,\s]$ is given in (\ref{As}).
Therefore, we get that the A-model is equivalent to the original model
\beq
Z_A=Z_O=\int D_Lg Dx e^{-S[g,x]}
\eeq{ZAZO}
(up to global issues \cite{GR,AABL} that we do not address here).

\section{The B-model}
\setcounter{equation}{0}

To integrate out the gauge field $A,\bA$ in (\ref{pint}) it is convenient to
re-express the non-local part of the action in terms of $A,\bA$ and the
conformal factor $\s$. The non-local part takes the form
\ber
S_{nonlocal}[A,\bA,\s]&=&\i ln(\det M(g) \det M(\bar{g}))\d\bd\s\nonumber\\
&=&{1\over 8\pi}(tr\ T_a)\int d^2z \Big({1\over \d}A^a+{1\over
\bd}\bA^a\Big)\sqrt{h}R^{(2)}\nonumber \\
&=&-{trT_a\over 2\pi}\int d^2z (A^a\bd \s + \bA^a\d\s).
\eer{As2}

The action $S[A,\bA,\l,x]+
S_{nonlocal}[A,\bA,\s]$ is bilinear in $A,\bA$ (it is
actually linear in $A$ and in $\bA$, separately). Therefore,
integrating out $A,\bA$ is simple and leads to
\ber
Z_B&=&\int D\l\ Dx\ DA\ D\bA\
e^{-(S[A,\bA,\l,x]+S_{nonlocal}[A,\bA,\s])}
\nonumber\\
&=&\int D\l\ Dx\ \det N(x,\l)\ e^{-S_B[\l,x,\s]},
\eer{eSB}
where the {\em dual} action $S_B$ is
\ber
S_B[\l,x,\s]&=&
\i \Big(
F_{ij}\d x^i\bd x^j - (\Phi-ln \det N)\d\bd\s \nonumber\\
&+&(\d\l_a-\d x^iF^{L}_{ia}+trT_a\d\s)
N^{ab}(\bd\l_b+F^{R}_{bj}\bd x^j-trT_b\bd\s) \Big)\, ,\nonumber\\ {}
\eer{Sdual}
and
\beq
N^{ab}(x,\l)=[(E(x)+\l_cf^c)^{-1}]^{ab} .
\eeq{Nxl}
Here the shift of the dilaton comes from a
Jacobian factor that arises from integrating over the gauge field
\cite{buscher} (see Appendix B), and the matrices $f^c$ have the structure
constants as components $(f^c)_{ab}=f^c_{ab}$.

Equation (\ref{Sdual}) is our key result: generally it is
non-local, since $\sigma$ appears explicitly in a form which does
not allow its replacement by the curvature,  without the use
of the inverse Laplacian. Since by construction (\ref{Sdual}) is
the form equivalent to model A, after the quantum corrections are
taken into account we can read off the non-local
corrections to the energy-momentum tensor needed to  make the
two models equivalent.
The variation with respect to $\sigma$ gives the trace of
the energy-momentum. Therefore, the explicit $\sigma$-dependent
terms in (\ref{Sdual}) give the difference at one-loop level
between the traces of the energy-momentum tensors for models A and B'
(recall that B' is the sigma-model part of the B model, without
the (non-local) dilaton-type correction). From the explicit
expressions we see that this difference will be
{\em always} a total derivative. This means, as discussed in the
introduction, that if model A was conformally invariant, then model B'
will have  at least a local, conserved dilation current.
Moreover, if model A is conformal invariant, then (\ref{Sdual})
is by construction $\sigma$-independent. Therefore, since the total
variation with respect to $\sigma$ is obtained by the $\beta$-function
equations in addition to the explicit $\sigma$ dependence,
we can directly obtain the $\beta$-functions of model B' by taking
(with negative sign) the $\sigma$-variations in (\ref{Sdual}).

We discuss now two typical examples.

\section{Example 1}
\setcounter{equation}{0}

In this section we present the first example. We discuss a two-dimensional
group $G_{JP}$ with generators $T_a$, $a=1,2$,
that we write in the adjoint representation as
\beq
T_2=P=\mat 0010, \qquad T_1=J=\mat 0001.
\eeq{TTPJ}
They obey the algebra
\beq
[J,P]=P \ \implies \ f^P_{JP}=-f^P_{PJ}=1,
\eeq{JPP}
and all other commutators and structure constants are 0.
The dual generators $\t{T}^a$ are given by
\beq
\t{T}^2=P^t=\mat 0100, \qquad \t{T}^1=J^t=J.
\eeq{PtJt}
It is easy to check that (\ref{trTT}) is satisfied.

We parametrize elements $u(\a,\b)\in G_{JP}$ by
\beq
u(\a,\b)=e^{\a P}e^{\b J}=\left(\matrix{1&0\cr \a&e^{\b}}\right).
\eeq{gab}
It is easy to check that $u(\a,\b)$ obey the group product
\beq
u(\a,\b)u(\a',\b')=u(\a+e^{\b}\a',\b+\b'),
\eeq{abab}
and the inverse is
\beq
\inv u(\a,\b) = u(-\a e^{-\b}, -\b)=\left(\matrix{1&0\cr
-\a e^{-\b}&e^{-\b}\cr}\right) .
\eeq{invgab}

To write $G_{JP}$ invariant actions it is convenient to parametrize $g\in
G_{JP}$ by
\beq
g(\p,\x)=\left(\matrix{1&0\cr \p&\inv{\x}\cr}\right) .
\eeq{gpx}
Then the isometry acts as
\beq
g(\p,\x)\to g'(\p',\x')=u(\a,\b)g(\p,\x)=\left(\matrix{1&0\cr
\p'&\inv{(\x')}\cr}\right) ,
\eeq{ugpx}
where
\beq
\p'=e^{\b}\p+\a, \qquad \x'=e^{-\b}\x.
\eeq{upux}
The $G_{JP}$ invariant elements in the action are constructed out of
\ber
(\inv g \d_{\mu}g)^P &\equiv& tr(P^t \inv g \d_{\mu} g)\ =\ \x\d_{\mu}\p,
\nonumber \\
(\inv g \d_{\mu}g)^J &\equiv& tr(J^t \inv g \d_{\mu} g)\ =\
-\inv{\x}\d_{\mu}\x.
\eer{gPtgJt}
For example, we will choose the action
\ber
S[\p,\x]&=&\i \Big((\inv g \d g)^P(\inv g \bd g)^J+(\inv g \d g)^J
(\inv g \bd g)^P\Big) \nonumber \\
&=& - \i (\d\p\bd\x+\d\x\bd\p).
\eer{Spx}
This action is of the form (\ref{noiso}) with $E_{PJ}=E_{JP}=1$, and all
other backgrounds are $0$. The background is a flat Minkowski space in two
dimensions and, therefore, the A-model is conformal.

To find the B-model we need the background matrix $N$ (\ref{Nxl}):
\beq
N^{ab}(x,\l)=[(E(x)+\l_cf^c)^{-1}]^{ab}={1\over 1-\l_P^2}
\left(\matrix{0&1+\l_P\cr 1-\l_P&0\cr}\right) ,
\eeq{NPJ}
Using
\beq
\det N = {1\over \l_P^2-1}\ , \qquad trT_a = (tr\ J)\delta_a^J = \delta_a^J,
\eeq{detNPJ}
one finds
\ber
S_B[\l]&=&\i \Big((\d\l_a+\delta_a^J \d\s)N^{ab}(\bd\l_b-\delta_b^J\bd\s)
+ln\ \det N\ \d\bd\s \Big) \nonumber\\
&=&\i \Big({1\over 1-\l_P^2}(\d\l_P\bd\l_J+\d\l_J\bd\l_P)
+{\rm total\ derivative}\Big), \nonumber\\
\eer{SBlPJ}
and
\beq
Z_B=\int {D\l_P D\l_J \over \l_P^2-1}\ e^{-S_B[\l]}.
\eeq{ZBPJ}
The background in $S_B$ is a 2-$d$ flat Minkowski space up to a conformal
factor:
\beq
ds^2=e^{\rho(\l_P)}(\d\l_P\bd\l_J+\d\l_J\bd\l_P), \qquad
\rho(\l_P)=-ln(1-\l_P^2),
\eeq{llflat}
and the curvature is
\beq
R=-4e^{-\rho(\l_P)}{\d\over \d_{\l_J}}{\d\over \d_{\l_P}}\rho(\l_J)=0.
\eeq{R0}
Therefore, the B-model is flat. Changing coordinates to
\beq
d\p={1\over \l_P^2-1}d\l_P, \qquad d\x=d\l_J,
\eeq{coch}
we get
\beq
Z_B=\int D\p\ D\x\ e^{-S[\p,\x]},\qquad S[\p,\x]=
- \i (\d\p\bd\x+\d\x\bd\p).
\eeq{ZBpx}
Therefore, the B-model (\ref{ZBPJ}) is equivalent to the A-model
(\ref{Spx}).

The fact that the B-model could be written in a local form, although the
anomaly was present, derives from the two-dimensional target-space nature.
In two dimensions any $\xi$ appearing in eq. (\ref{bxx}) is of the type
(\ref{zeta}). Therefore, the energy-momentum tensor is improvable by a
dilaton term. Indeed, in the example above, the  term induced by the
anomaly is of a dilaton type (up to a total derivative), and moreover, it
exactly cancels the usual dilaton term arising by duality.

\section{Example 2}
\setcounter{equation}{0}

In this section we present the four-dimensional Bianchi V
cosmological example considered in ref. \cite{GRV}.
We discuss a three-dimensional group $G_V$ with generators $T_a$,
$a=1,2,3$, that we write in the adjoint representation as
\beq
T_1=\left(\matrix{0&0&0\cr 0&1&0\cr 0&0&1\cr}\right),\qquad
T_2=\left(\matrix{0&0&0\cr 1&0&0\cr 0&0&0\cr}\right),\qquad
T_3=\left(\matrix{0&0&0\cr 0&0&0\cr 1&0&0\cr}\right).
\eeq{T123}
They obey the algebra
\ber
[T_1,T_2]=T_2,\qquad [T_1,T_3]=T_3,\qquad [T_2,T_3]=0\ \implies\nonumber \\
f_{12}^2=f_{13}^3=1=-f_{21}^2=-f_{31}^3, \qquad f_{ab}^c=0\;\;\; {\rm
otherwise}.
\eer{GValg}
The dual generators $\t{T}^a$ are given by
\beq
\t{T}^1={1\over 2}T_1,\qquad \t{T}^2=(T_2)^t,\qquad \t{T}^3=(T_3)^t.
\eeq{tTtt}
It is easy to check that (\ref{trTT}) is satisfied.

We parametrize elements $u(\a,\vb)\in G_V$, $\vb=(\b_1,\b_2)$ by
\beq
u(\a,\vb)=e^{\b_1T_2+\b_2T_3}e^{\a T_1}=
\left(\matrix{1&0&0\cr \b_1&e^{\a}&0\cr \b_2&0&e^{\a}\cr}\right).
\eeq{uavb}
It is easy to check that $u(\a,\vb)$ obey the group product
\beq
u(\a,\vb)u(\a',\vb')=u(\a+\a',\vb+e^{\a}\vb'),
\eeq{avbavb}
and the inverse is
\beq
\inv u(\a,\vb) = u(-\a, -\vb e^{-\a})=
\left(\matrix{1&0&0\cr -\b_1e^{-\a}&e^{-\a}&0\cr -\b_2e^{-\a}&0&e^{-\a}\cr}
\right).
\eeq{invuab}

Let $g(\p,\vx)\in G_V$, $\vx=(\x_1,\x_2)$ be
\beq
g(\p,\vx)=\left(\matrix{1&0&0\cr \x_1&e^{\p}&0\cr \x_2&0&e^{\p}\cr}\right).
\eeq{upvx}
Then the isometry $u$ acts on $g$ by a left multiplication
\beq
g(\p,\vx)\to
g'(\p',\vx')=u(\a,\vb)g(\p,\vx)=g(\p+\a,e^{\a}\vx+\vb).
\eeq{isogpvx}
The $G_V$ invariant elements in the action are constructed out of
\ber
(\inv g \d_{\mu}g)^1 &\equiv& tr(\t{T}^1 \inv g \d_{\mu} g)\ =\d_{\mu}\p,
\nonumber \\
(\inv g \d_{\mu}g)^2 &\equiv& tr(T^{2t} \inv g \d_{\mu} g)\ =
\ e^{-\p}\d_{\mu}\x_1, \nonumber\\
(\inv g \d_{\mu}g)^3 &\equiv& tr(T^{3t} \inv g \d_{\mu} g)\ =
\ e^{-\p}\d_{\mu}\x_2.
\eer{g123g}

To write the four-dimensional Bianchi V model we choose to insert in eq.
(\ref{noiso}):
\beq
x^i\equiv \{ t \},\qquad E_{ab}(t)=a(t)^2\delta_{ab},\qquad F(t)=-1,\qquad
F^L=F^R=\Phi=0.
\eeq{Eat}
The action takes the form
\ber
S[\p,\vx,t]&=&\i \Big(a(t)^2\delta_{ab}(\inv g \d g)^a(\inv g \bd g)^b
-\d t\bd t\Big)\nonumber\\
&=& \i \Big(-\d t \bd t + a(t)^2[\d\p\bd
\p+e^{-2\p}(\d\x_1\bd\x_1+\d\x_2\bd\x_2)]\Big).
\nonumber \\
\eer{Spxt}
Conformal invariance requires
\beq
a(t)=t.
\eeq{att}
In this case the metric is flat (it is the interior of the light-cone in
Minkowski space).
In the following we consider the conformal model (\ref{Spxt}),(\ref{att}).

To find the B-model we need the background matrix $N$ (\ref{Nxl}):
\beq
N^{ab}(t,\l)=[(E(t)+\l_cf^c)^{-1}]^{ab}=
{1\over t^2(t^4+\l_2^2+\l_3^2)}
\left(\matrix{t^4&-\l_2t^2&-\l_3t^2\cr \l_2t^2&t^4+\l_3^2&-\l_2\l_3\cr
\l_3t^2&-\l_2\l_3&t^4+\l_2^2\cr}\right) ,
\eeq{Nllt}
Using
\beq
\det N = {1\over t^2(t^4+\l_2^2+\l_3^2)}\ , \qquad
tr T_a = (tr\ T_1)\delta_a^1 = 2\delta_a^1,
\eeq{detNttt}
one finds
\ber
S_B[\l,t]&=&\i \Big((\d\l_a+2\delta_a^1 \d\s)N^{ab}(\bd\l_b-2\delta_b^1\bd\s)
\nonumber\\&-&\d t\bd t+ln\ \det N\ \d\bd\s \Big) \nonumber\\
&=& S_0[\l,t]+S_1[\l,t]+S_2[\l,t],
\eer{SBV}
\beq
S_0[\l,t]=\i \Big(\d\l_a N^{ab}(t,\l) \bd\l_b + ln\ \det N \d\bd\s\Big) ,
\eeq{S0lt}
\beq
S_1[\l,t]={2\over 2\pi}\int d^2z\
\s\Big(\bd(N^{a1}(t,\l)\d\l_a)-\d(N^{1b}(t,\l)\bd\l_b)\Big).
\eeq{S1}
\beq
S_2[\l,t]=-{2^2\over 2\pi}\int d^2z\ N^{11}(t,\l)\d\s\bd\s.
\eeq{S2}

The A-model is conformal and, therefore, one expects the B-model to be
conformal, {\it i.e.},
\beq
\pi{\delta S_B\over \delta\s}=T_{z\bar{z}}=0=
\pi{\delta (S_0+S_1+S_2)\over \delta\s}=
T^0_{z\bar{z}}+T^1_{z\bar{z}}+T^2_{z\bar{z}}.
\eeq{dSds}
Variation of $S_0$ with respect to the conformal factor $\s$ gives
\beq
\pi{\delta S_0\over \delta\s}=T^0_{z\bar{z}}=
{1\over 2}[\b_{G_{IJ}}(X)+\b_{B_{IJ}}(X)]\d X^I \bd X^J +
{1\over 2}\b_{\Phi}(X)\d\bd\s,
\eeq{T0}
where
\ber
X^I\equiv\{ \l^a,t\}, \qquad \Phi=-ln\ \det N, \nonumber\\
G_{ab}={1\over 2}(N+N^t)^{ab}, \quad G_{tt}=-1, \qquad {\rm
otherwise}\;\; G_{IJ}=0, \nonumber\\
B_{ab}= {1\over 2}(N-N^t)^{ab}, \qquad {\rm otherwise}\;\; B_{IJ}=0,
\eer{XGB}
and to one-loop order the beta-functions are \cite{bfun}
\ber
\b_{G_{IJ}}&=&R_{IJ}-{1\over 4}H^2_{IJ}-\n_I\n_J\Phi , \nonumber\\
\b_{B_{IJ}}&=&-{1\over 2}\n^KH_{IJK}-{1\over 2} \n^K\Phi H_{IJK} , \nonumber\\
\b_{\Phi}&=&R-{1\over 12}H^2-2\n^2\Phi-(\n\Phi)^2-{2(c-4)\over 3} ,
\eer{beta}
where
\beq
H_{IJK}=\n_{[I}B_{JK]}, \qquad H^2_{IJ}=H_{IKL}H_J^{KL}.
\eeq{HH2}

Variation of $S_1$ and $S_2$ with respect to $\s$ gives
\ber
\pi{\delta S_1\over \delta\s}&=&T^1_{z\bar{z}}\ =\
\bd(N^{a1}(t,\l)\d\l_a)-\d(N^{1b}(t,\l)\bd\l_b)+\cal{O}(\s),
\nonumber\\
\pi{\delta S_2\over \delta\s}&=&T^2_{z\bar{z}}\ =\ 0+\cal{O}(\s).
\eer{dS1ds}
Here the leading order comes from naive variation with respect to $\s$;
higher order corrections in $\s$ require more careful regularization of
(\ref{S1}),(\ref{S2}), as done in deriving $T_{z\bar{z}}^0$ (\ref{T0}).
We should note that the leading order contribution to
$T^1_{z\bar{z}}+T^2_{z\bar{z}}$ is a total derivative.
As will be shown below, using the equation of motion
\beq
\d(N^{ab}\bd\l_b)+\bd(N^{ba}\d\l_b)
-{\delta N^{bc} \over \delta\l_a}\d\l_b\bd\l_c
=0+\cal{O}(\s),
\eeq{eleq}
one finds that
\beq
T^0_{z\bar{z}}+T^1_{z\bar{z}}+T^2_{z\bar{z}}=0+\cal{O}(\s).
\eeq{TTT0}

In the following we present the detailed calculations.
The $\s$-model described by the action $S_0$ (\ref{S0lt}) has non-zero
$\b$-functions \cite{GRV}. Explicitly, we have the metric
\beq
ds^2=-dt^2+{t^2\over 4x(t^4+x)}dx^2+{x\over t^2}dy^2+{t^2\over t^4+x}dz^2,
\eeq{metric}
the antisymmetric tensor
\beq
B={1\over 2(t^4+x)}dx\wedge dz,
\eeq{asymm}
and the dilaton
\beq
\Phi=ln\ t^2 + ln(t^4+x).
\eeq{dilaton}
These are expressed in terms of the variables $x^{\mu}=\{t,x,y,z\}$
defined by:
\beq
\l_1=z,\qquad \l_2=\sqrt{x}\cos y, \qquad \l_3=\sqrt{x}\sin y.
\eeq{xyz}
In our case, we have for the Ricci tensor the following non-vanishing
components:
\ber
R_{tt}&=&{-6t^8+16t^4x-2x^2\over t^2(t^4+x)^2}, \qquad
R_{tx}\ =\ {-t^4+x\over t(t^4+x)^2}, \nonumber\\
R_{xx}&=&{2t^8-7t^4x-x^2\over 2x(t^4+x)^3},\qquad
R_{yy}\ =\ {4x\over t^4+x},\nonumber\\
R_{zz}&=&{6t^8-10t^4x\over (t^4+x)^3}.
\eer{RRR}
For $H_{\mu\nu}^2$ we get the non-vanishing components
\beq
(H^2)_{tt}={32t^2x\over (t^4+x)^2}, \qquad
(H^2)_{xx}={-8t^4\over (t^4+x)^3}, \qquad
(H^2)_{zz}={-32t^4x\over (t^4+x)^3}.
\eeq{HHH}
Including the the $\n_{\mu}\n_{\nu}\Phi$ contributions,
\ber
\n_{t}\n_{t}\Phi&=&{2(-3t^8+4t^4x-x^2)\over t^2(t^4+x)^2},\qquad
\n_{t}\n_{x}\Phi\ =\ {-(3t^4+x)\over t(t^4+x)^2}, \nonumber \\
\n_{x}\n_{x}\Phi &=& {4t^8-t^4x-x^2\over 2x(t^4+x)^3}, \nonumber \\
\n_{y}\n_{y}\Phi&=&{4x(2t^4+x)\over t^4(t^4+x)}, \qquad
\n_{z}\n_{z}\Phi\ =\ {2(3t^8-3t^4x-2x^2)\over (t^4+x)^3},
\eer{ddPhi}
we have for
$\b_{G_{\mu\nu}}$ in (\ref{beta}):
\ber
\b_{G_{tx}} &=& {2\over t(t^4+x)}, \qquad
\b_{G_{xx}}\ =\ {-t^4\over x(t^4+x)^2}, \qquad
\b_{G_{yy}}\ =\ {-4x\over t^4}, \nonumber\\
\b_{G_{zz}} &=& {4x\over (t^4+x)^2},
\eer{betaG}
and for $\b_{B_{\mu\nu}}$ in (\ref{beta}):
\ber
\b_{B_{\mu\nu}}&=&-{1\over 2}\n^\rho H_{\mu\nu\rho}-{1\over 2}
\n^\rho \Phi H_{\mu\nu\rho} \nonumber\\
\b_{B_{tz}}&=&{-4t^5\over (t^4+x)^2}-{4tx\over (t^4+x)^2}\ =\
{-4t\over t^4+x},\nonumber\\
\b_{B_{xz}} &=& {4t^6\over (t^4+x)^3}-{2t^2(3t^4+x)\over (t^4+x)^3}\ =\
{-2t^2\over (t^4+x)^2}.
\eer{betaB}
All other components of $\b_G$ and $\b_B$ are zero.
The dilaton $\b$-function in (\ref{beta}) turns out to be a constant:
\ber
\b_{\Phi}&=&R-{1\over 12}H^2-[2\n^2\Phi-(\n\Phi)^2]-{2(c-4)\over 3}
\nonumber\\
&=&{4t^4(5t^4-9x)\over t^2(t^4+x)^2}-{8t^2x\over (t^4+x)^2}
-{4t^2(5t^4-7x)\over (t^4+x)^2}-{2(c-4)\over 3}\nonumber\\
&=&{2(4-c)\over 3},
\eer{betaP}
and therefore, $\b_{\Phi}$ vanishes if the central charge is $c=4$, as for
the A-model~\footnote{
Note that here $\b_G$ and $\b_B$ do not vanish and, therefore, the fact that
$\b_{\Phi}$ is a constant (actually zero) is not obvious.
Even if the dilaton beta-function would
not vanish, there could be $\cal{O}(\s)$ contributions in (\ref{dS1ds}) that
would cancel $\b_{\Phi}\d\bd\s$ in $T_{z\bar{z}}$. In our case we conclude
that such $\cal{O}(\s)$ corrections must vanish.}.

Although we have found non-vanishing contributions to the $\b$-functions,
the additional non-local term (\ref{S1}), resulting from the anomaly
(\ref{As}), has an extra explicit $\s$ dependence which should be taken
into account. The variation of this term with respect to $\s$ exactly
cancels the above $\b$-functions, recovering the conformal symmetry of the
original model. Indeed, from (\ref{S1}) one finds
\ber
\delta_{\s} S_1 &=& {1\over \pi}\int d^2z\ \delta\s
\Big[\bd\Big({1\over t^4+x}(t^2\d
z+{1\over 2}\d x)\Big)-\d\Big({1\over t^4+x}(t^2\bd z-{1\over 2}\bd
x)\Big)\Big]\nonumber\\
&=& {1\over \pi}\int d^2z\ \delta\s
\Big[{2t(-t^4+x)\over (t^4+x)^2}(\bd t\d z - \d t \bd z ) -
{t^2\over (t^4+x)^2}(\bd x\d z - \d x \bd z ) \nonumber \\
&-& {1\over (t^4+x)^2} [2t^3(\bd t\d x+\d t\bd x)+\bd x\d x]+{1\over
t^4+x}\d\bd x\Big].
\eer{dS1}
Substituting the equation of motion,
\ber
\d\bd x&=&{t^4+2x\over 2x(t^4+x)}\d x\bd x\ +\ {t^4-x\over t(t^4+x)}(\d t\bd
x+\bd t\d x)\nonumber\\
&+& {2x(t^4+x)\over t^4 }\d y\bd y\ -\ {2x\over t^4+x}\d z \bd z\ +\
{4tx\over t^4+x}(\d t\bd z-\bd t \d z), \nonumber\\
\eer{eqofmo}
into eq. (\ref{dS1}), one finds
\beq
\delta_{\s} S_1=-\i [(\b_{G_{\mu\nu}}+\b_{B_{\mu\nu}})\d x^{\mu}\bd
x^{\nu}]\delta\s ,
\eeq{findS1}
with the same $\b_G$ and $\b_B$ of eqs. (\ref{betaG}), (\ref{betaB}).
Therefore, the total action $S_0+S_1$ is $\s$ independent, as it should be.

Finally, let us emphasize
that we have checked cancellation only to leading order in $\s$,
namely, for $\s=0$.
However, we should note that since the $\cal{O}(\s)$ corrections in
(\ref{TTT0}) must vanish, this can be used to find the $\cal{O}(\s)$
contribution to the variation of the non-local action with respect to
the conformal factor: $\delta(S_1+S_2)/\delta\s$.

To appreciate the possible effect of the mixed anomaly term on the actual
value of the Virasoro central charge, consider the
extremely simple system of a single scalar with a linear dilaton term:
\beq
L_A=\d x\bd x+Q \sqrt{h} R^{(2)}x.
\eeq{LAx}
This is a conformal system with central charge $c=1+3Q^2$. Applying the
above procedure we get in the new variables $A$ and $\l$:
\beq
L=A\bA+\l F+Q(\d\bA+\bd A){1\over \Box}  \sqrt{h} R^{(2)},
\eeq{LAl}
where now the non-local term does not result from a Jacobian anomaly but
rather from the explicit $Q$ term in the original action. Integrating out
$A$ will now give:
\beq
L_B=\d\l\bd\l+{1\over 2}Q^2 \sqrt{h}R^{(2)}{1\over \Box}\sqrt{h}R^{(2)}.
\eeq{LBl}
The scalar $\l$ is now a
normal massless scalar field with $c=1$, and the difference between the
original central charge and the final one is compensated by the explicit
non-local $Q^2$ term in $L_B$.
Note that this is an order $\s^2$ term (exact, in this example) as expected.

\section{The relation between local energy-momentum tensors in models A and B}
\setcounter{equation}{0}

In this work we considered the two-dimensional quantum field theory
(\ref{ZlAA}). For the purpose of the discussion in this section,
let us neglect the
determinants term $\exp {J(A,\bA)}$ in (\ref{ZlAA}), and consider the theory
\beq
\int D\l\ Dx\ DA\ D\bA\ e^{-S[A,\bA,\l,x]},
\eeq{I}
where $S[A,\bA,\l,x]$ is given in (\ref{SxA}). In a flat worldsheet
background the two transformations of (\ref{I}) leading to models A and B
are valid for any isometry group $G$.
The field theories described by A and B are
equivalent, {\em i.e.}, there exists
an exact mapping between the Hilbert spaces of the
two theories which preserves the spectrum.
The correspondence between the various local operators
in the two theories is more involved.
In particular, the relation between local energy-momentum tensors, and
therefore, the coupling to a non-flat background in the two models, is the
subject of this section.

We will use two methods which should be equivalent:

\noindent
a) Duality transformation in a curved worldsheet background.

\noindent
b) Ward identities involving the two energy-momentum tensors calculated with
the action in a flat background.

Consider (\ref{I}) in a curved background. In the conformal gauge,
classically the conformal factor does not appear. Through the measure,
however, a dependence on the conformal factor may appear. This was
discussed in section 2. It was shown that in a curved background the theory
(\ref{I})
does not lead to model A. If we want to get exactly model A we should add
to (\ref{I}) the anomalous piece (\ref{As}).\footnote{In section 2, this term
was included in the theory from the beginning, since we were careful to keep
the term $\exp J$ in the measure of Z (\ref{ZlAA}).}

Equivalently, by taking functional derivatives with respect to the
worldsheet metric
and gauge field, we conclude that the energy-momentum tensor of model A
should have an anomalous correlator with the current coupled to the gauge
field when calculated with the flat background action.

As was remarked in section 2, for the ``pure'' Weyl anomaly,
{\em i.e.}, the contributions to the term
$\sqrt{h}R^{(2)}{1\over\Box}R^{(2)}$ in the effective action, there is no
multiplicative anomaly. As a consequence we expect that once the current
anomaly $\n A{1\over\Box}R^{(2)}$ is cancelled by an  appropriate
counter term, the A- and B-models, if conformal, will
have the same central charge.

We present now an independent calculation, confirming the conclusions above.

We consider for action (\ref{I}) the topologically conserved
current~\footnote{The conservation equation of this current is the
``dual Bianchi identity'' presented in ref. \cite{GR}.}
\beq
J_{a\a}=\e_{\a\b}\d^{\b}\l_a .
\eeq{topJ}
In the following, we choose ``$a$'' to be a quasi-Abelian direction.
Using the equations of motion, $J_{a\a}$ is given by
\beq
J_{a\a}={\delta S\over \delta A^{a\a}}+\e_{\a\b}f^b_{ac}A^{c\b}\l_b.
\eeq{emJ}
We want to express the topological current of model B, $J_{a\a}$, in terms
of the variables of model A. The variables $\l_b$ can be calculated by
adding to $S$ a source term, $\l_b K^b$, and taking the derivative of $S$
with respect to the source:
\beq
\l_a(\z)={\delta S\over\delta K^a(\z)}=\int d\z'{\delta S\over\delta
A_{\a}^b(\z')}{\delta A_{\a}^b(\z')\over\delta K^a(\z)}.
\eeq{lK}
To first order in $K$, the gauge field is given by:
\beq
A_{\a}^a(\z)=(g^{-1}\d_{\a}g)^a+\Big[h^{-1}(\z){1\over\Box_{\z\z'}}\e_{\a\b}
\d^{\b} (gKg^{-1})_{\z'}g(\z)\Big]^a ,
\eeq{AgK}
where $g^{-1}\d_{\a}g=A_{\a}$.
Using (\ref{AgK}) the final result for the current is:
\ber
J_{a\a}(\z)&=&{\delta S\over \delta A^{a\a}(\z)}
+\e_{\a\b}f^b_{ac}A^{c\b}(\z)\ \times
\nonumber\\ &\times&
M_b^{\ d}(g^{-1}(\z)){1\over\Box_{\z\z'}}\e_{\g\delta}\d^{\delta'}\Big[
{\delta S\over\delta A^{\g e}(\z')}M_d^{\ e}(g(\z'))\Big].
\eer{JMM}
Here $M(g)$ is the adjoint representation (\ref{Madj}).
If $S$ has a quadratic dependence on $A^{a\a}$, the second term in
(\ref{JMM}) contains
\beq
\e^{\g\delta}\d_{\delta}(gg^{-1}\d_{\g}g g^{-1})
=\e^{\g\delta}[g^{-1}\d_{\g}g,g^{-1}\d_{\delta}g],
\eeq{egggg}
which does not have a component in a quasi-Abelian direction. It follows
that up to terms cubic in the fields,
\beq
J_{a\a}(\z)={\delta S\over \delta A^{a\a}(\z)}.
\eeq{utct}
In particular, for the model based on the group $G_V$ in section 6,
$J_{a\a}$ has the form:
\beq
J_{\a}=E_{11}(t)\d_{\a}\p.
\eeq{Jphi}
Considering at one-loop level the correlator with the energy-momentum
tensor we obtain a non-zero, anomalous contribution from the term
\beq
T_{\a\b}=E_{11}(\d_{\a}\x_1\d_{\b}\x_1+\d_{\a}\x_2\d_{\b}\x_2)
\eeq{TExx}
in the presence of the interaction term
\beq
L_{int}=-2E_{11}\p(\d_{\a}\x_1\d^{\a}\x_1+\d_{\a}\x_2\d^{\a}\x_2).
\eeq{Lint}
On the other hand, if we calculate directly the correlator in model B,
since the current (\ref{topJ}) is topological in terms of the new local
field $\l_a$, it cannot be anomalous: the derivative and $\e$-tensors can
be taken outside the correlator for which dimensional regularization can be
used. We conclude that the energy-momentum tensors defined by (\ref{I}) for
models A and B differ by their correlators to the topological current
(\ref{topJ}). This current is the source of the gauge field
$g^{-1}\d_{\a}g$ and, therefore, we reproduce this way the anomaly
contribution in eq. (\ref{As}).

\section{On localizing the mixed anomaly term}
\setcounter{equation}{0}

Let us write down the formal equality
\ber
&{}&\int Ds\ Dt\ Dx\ e^{-\i \Big(S_0[x]+2\d s\bd t+Q(x){1\over \Box}
\sqrt{h}R^{(2)} \Big)} \nonumber\\
&=&\int Du\ Dv\ Dx\
e^{-\i \Big(S_0[x]+2\d u\bd v+4vQ(x)+{1\over 4} u \sqrt{h}R^{(2)}\Big)},
\eer{uv1}
where $\Box=2\d\bd$. (The models are not necessarily the same.)
One can show this equality by integrating out $s,t$ on the
left-hand side, and comparing with the integration over $u,v$ on the
right-hand side of eq. (\ref{uv1}).
Some choices of $Q(x)$ are interesting. For example, if $S_0[x]$ is a sigma
model with beta-functions $\b_G$, $\b_B$, then with the choice
\beq
Q(x)=-{1\over 2}(\b_{G_{\mu\nu}}+\b_{B_{\mu\nu}})\d x^{\mu}\bd x^{\nu},
\eeq{uv2}
the model $S_0[x]+2\d s\bd t+Q(x){1\over \Box}\sqrt{h}R^{(2)}$ is manifestly
independent of $\s$, to leading order, which implies that the model
$S_0[x]+2\d u\bd v+4vQ(x)+{1\over 4} u \sqrt{h}R^{(2)}$ is conformally
invariant to leading order. (For a discussion about higher order corrections,
including a dilaton beta-function in (\ref{uv2}), see section 6.)
Similar models were discussed in ref.~\cite{arkady}.

Moreover, if $S_0=S[A,\bA,\l,x]$ (\ref{SxA}), and if we choose
\beq
Q(A,\bA)={1\over 4}(tr T_a)(\d \bA^a+\bd A^a),
\eeq{uv3}
then, using eq. (\ref{uv1}), one finds that
\ber
&{}&\int Dx\ D\l\ DA\ D\bA\ Ds\ Dt\ e^{-\i S_a[x,A,\l,s,t]}
\nonumber\\ &=&
\int Dx\ D\l\ DA\ D\bA\ Du\ Dv\ e^{-\i S_b[x,A,\l,u,v]},
\eer{uv4}
where
\ber
S_a&=&S[A,\bA,\l,x]+S_{nonlocal}[A,\bA,\s]+\i\ 2\d s\bd t, \nonumber \\
S_b&=&S[A,\bA,\l,x]\nonumber\\
&+&\i \Big( 2\d u\bd v+v(tr T_a)(\d \bA^a+\bd A^a)+
{1\over 4} u \sqrt{h}R^{(2)}\Big).
\eer{uv5}
Here $S[A,\bA,\l,x]$ and $S_{nonlocal}[A,\bA,\s]$ are given in eqs.
(\ref{SxA}) and (\ref{As2}), respectively.
Now, integrating out $\l$ on the left-hand side of eq. (\ref{uv4}), we recover
the A-model in section 3, with an additional decoupled null kinetic term
$\d s\bd t$. On the other hand, integrating out $A,\bA$ on the right-hand side
of eq. (\ref{uv4}), we get a {\em local} dual theory with action
\ber
S_b[\l,x,u,v]&=&
\i \Big(  2\d u\bd v - (\Phi-u-ln \det N)\d\bd\s \nonumber\\
&+&(\d\l_a-\d x^iF^{L}_{ia}+trT_a\d v)
N^{ab}(\bd\l_b+F^{R}_{bj}\bd x^j-trT_b\bd v)\nonumber\\
&+& F_{ij}\d x^i\bd x^j \Big)\, .
\eer{uv6}
Comparing to $S_B[\l,x,\s]$ in eq. (\ref{Sdual}), we see that there is an
additional null kinetic term, $\d u \bd v$,  the term leading
to non-locality, $trT_a\d\s$, has been replaced by $trT_a\d v$, and
the dilaton $\Phi$ is shifted to $\Phi-u$. Therefore, the additional null
coordinates $u,v$ provide a localization of the B-model in a higher dimensional
sigma model.

Finally, we should mention that another way to localize the mixed anomaly
term, within a larger sigma model framework, is to bosonize the ghosts in eq.
(\ref{bcbc}).
By integrating out the gauge field in (\ref{ZlAA}), one
will obtain a local sigma model action.

\vskip .3in \noindent
{\bf Acknowledgements} \vskip .2in \noindent
We would like to thank R. Ricci and M. Ro\v cek for discussions, and E.
Verlinde for a discussion concerning the local representation. We also thank
O. Fonarev and Z. Ligeti  for help with ``Mathematica".
SE, AG and ER wish to thank the Theory Division at CERN for its warm
hospitality.
SE and ER wish to thank the Einstein center.
The work of SE is supported in part by the BRF - the Basic Research
Foundation.
The work of AG is supported in part by BSF - American-Israel Bi-National
Science Foundation and by an Alon fellowship.
The work of ER is supported in part by BSF and by the BRF.
The work of AS is supported in part by BSF.

\section{Appendix A}
\setcounter{equation}{0}

In this Appendix we prove the second equality in (\ref{As}), namely
\beq
ln(\det M(g)\det M(\bar{g}))=
tr T_a \Big({1\over \d}A^a+{1\over \bd}\bA^a\Big).
\eeq{prove}
If all the generators are traceless, there is nothing to prove.
Suppose not all of them are traceless.
We choose a basis for the generators $T_a$ such that
\beq
tr\ T_1 \neq 0\ , \qquad tr\ T_a=0\ , \qquad a=2,...,D={\dim}\ G\ .
\eeq{trTa}
This is always possible by redefining $T_a\to T_a-(tr\ T_a/tr\ T_1)T_1$.
Therefore, we want to prove that
\beq
ln\ \det\ M(g) = (tr\ T_1){1\over \d}A^1.
\eeq{MA1}
Recall that $A=\inv g \d g$, and we choose to parametrize $g$ by
\beq
g(\a_a)=g_2g_1\ , \qquad g_1=e^{\a^1 T_1}\ , \quad
g_2=e^{\sum_{a=2}^D\a^a T_a}\ .
\eeq{h2h1}
Recall also that $M_a^{\ b}(g)T_b=gT_a\inv g\ \implies\
M_a^{\ b}(g)tr\ T_b = tr(gT_a\inv g) = tr\ T_a$, which implies
\beq
M_1^{\ 1}(g)=1\ , \qquad M_a^{\ 1}(g)=0\quad{\rm for}\quad a\neq 1.
\eeq{Ma11}

Let us now find $A^1$ in terms of $\a$:
\ber
A^1&=&tr(A\t{T}^1)\ \ =\ \ tr(\inv g \d g \t{T}^1)\nonumber\\
&=&tr(\inv{g_1}\d g_1 \t{T}^1)+tr(\inv{g_1}\inv{g_2}\d g_2 g_1\t{T}^1)
\nonumber\\
&=&\d\a^1\ tr(T_1\t{T}^1)+(\inv{g_2}\d g_2)^a tr(\inv{g_1}T_a g_1 \t{T}^1)
\nonumber\\
&=&\d\a^1+(\inv{g_2}\d g_2)^a M_a^{\ 1}(\inv{g_1})=\d\a^1 .
\eer{A1da1}
Here we have used eqs. (\ref{trTT}), (\ref{Madj}), (\ref{h2h1}), and
in the last equality we have used eq. (\ref{Ma11}).
Moreover,
\beq
ln\ \det\ M(g)=ln\ \det\ M(g_1) + ln\ \det\ M(g_2) = tr\ ln\ M(g_1) = tr\ T_1
\a^1\ .
\eeq{T1a1}
Here we have used eq. (\ref{h2h1}), and the fact that $tr\ T_a=0$ for
$a\neq 1$ implies $\det\ M(g_2)=1$.
Inserting $\a_1=(1/\d)A^1$ (\ref{A1da1}) in (\ref{T1a1}) proves eq.
(\ref{MA1}).

\section{Appendix B}
\setcounter{equation}{0}

In this Appendix we study the conformal factor-dependent terms which may
appear when Gaussian integration over the $A$-variables is performed in
(\ref{eSB}). The origin of these terms is the implicit dependence on the
conformal factor in the measure of $A$ \cite{buscher}. This can be made
explicit by expressing the vector field $A^a,\bA^b$ in terms of scalar
variables $y^a,\by^{\bb}$:
\beq
A^a\equiv \d y^a, \qquad  \bA^b \equiv \bd\by^{\bb}.
\eeq{AAyy}
Since the $x$ and $\l$ variables do not participate in this effect, and the
linear terms in $A$ do not contribute (we can complete the square in
$A,\bA$), in order to find the $\s$-dependence of the $A$ measure (to
one loop), we can replace $S[A,\bA,\l,x]+S_{nonlocal}[A,\bA]$ in
(\ref{eSB}),(\ref{SxA}),(\ref{As}) by the action
\beq
S[y,\by,x]=\i [E_{ab}(x)\d y^a\bd\by^{\bb}+\d x \bd x],
\eeq{Syx}
where $x$ is a single field with flat metric.
The variation with respect to the conformal factor in the $A$-measure is
obtained by calculating the trace of the energy-momentum tensor of
(\ref{Syx}). This is given by the general beta-function equations
\cite{bfun},
\beq
T_{z\bar{z}}={1\over 2}(R_{IJ}-{1\over 4}H^2_{IJ}-{1\over 2}\n^KH_{IJK})\d
X^I \bd X^J,
\eeq{Txx}
where $X^I$ in our case include $x,y^a,\by^{\bb}$, and $R,H$ are the Ricci
curvature and antisymmetric field strength calculated from the metric and
torsion. From eq. (\ref{Syx}) we identify the metric $G$:
\beq
G_{xx}=1,\qquad G_{a\bb}=G_{\bb a}={E_{ab}\over 2},
\eeq{GE}
and torsion $B$:
\beq
B_{a\bb}=-B_{\bb a}={E_{ab}\over 2},
\eeq{BE}
all other components being zero. Using the special (quasi-K\"ahlerian) form
of $G$ and $B$ and their independence of $y^a,\by^{\bb}$, eq. (\ref{Txx})
is easily calculated:
\ber
2T_{z\bar{z}}&=&
{\d^2 ln\sqrt{\det G}\over \d x^2}\d x\bd x+{1\over 2}{\d\over \d x}
(ln\sqrt{\det G}){\d E_{ab}\over \d x}\d y^a \bd \by^{\bb}\nonumber\\
&=&\d\bd(ln\sqrt{\det G})\ =\ \d\bd (ln\ \det E),
\eer{2T}
where in the second equality we used the equation of motion, and in the
last step we used the special form of $G$ (\ref{GE}).
To get eq. (\ref{2T}) it is convenient to use:
\ber
R_{ij}&=&{\d^2 ln\sqrt{g}\over \d x^i\d x^j} - {\d\over \d x^k} \Gamma_{ij}^k
+\Gamma_{ik}^m\Gamma_{mj}^k-\Gamma_{ij}^m{\d ln\sqrt{g}\over\d x^m},
\nonumber\\
 g^{ij}{ {\d g_{ij}}\over{ \d x^m}} &=& 2{ { \d  ln\sqrt{g}}\over{\d x^m}}
\eer{RGamma}

Therefore, the conformal factor dependence produces a dilaton field
$\Phi(x)$ for the $x$ variables:
\beq
\Phi(x)=ln\ \det E.
\eeq{PE}
Since the result in (\ref{2T}) is given in terms of the worldsheet
variables, the result clearly is not dependent on the number of $x$
variables.

\end{document}